# Cometal addition effect on superconducting properties and granular behaviours of polycrystalline FeSe$_{0.5}$Te$_{0.5}$

Manasa Manasa[1], Mohammad Azam[1], Tatiana Zajarniuk[2], Ryszard Diduszko[2,3], Tomasz Cetner[1], Andrzej Morawski[1], Andrzej Wiśniewski[2], Shiv J. Singh[1,*]

[1] Institute of High Pressure Physics (IHPP), Polish Academy of Sciences, Sokolowska 29/37, Warsaw, Poland
[2] Institute of Physics, Polish Academy of Sciences, aleja Lotników 32/46, 02-668 Warsaw, Poland
[3] Institute of Microelectronics and Photonics, Wólczyńska 133, 01-919 Warsaw, Poland

* Correspondence: sjs@unipress.waw.pl

**Abstract:** The enhanced performance of superconducting FeSe$_{0.5}$Te$_{0.5}$ materials with added micro-sized Pb and Sn particles is presented. A series of Pb and Sn added FeSe$_{0.5}$Te$_{0.5}$ (FeSe$_{0.5}$Te$_{0.5}$ + $x$Pb + $y$Sn; $x = y$ = 0-0.1) bulks are fabricated by solid-state reaction method and characterized through various measurements. A very small amount of Sn and Pb additions ($x = y \leq 0.02$) enhance the transition temperature ($T_c^{onset}$) of pure FeSe$_{0.5}$Te$_{0.5}$ by ~1 K, sharpening the superconducting transition and improving the metallic nature in the normal state, whereas larger metal additions ($x = y \geq 0.03$) reduce $T_c^{onset}$ by broadening the superconducting transition. Microstructural analysis and transport studies suggest that at $x=y>0.02$, Pb and Sn additions enhance the impurity phases, reduce the coupling between grains, and suppress the superconducting percolation, leading to a broad transition. FeSe$_{0.5}$Te$_{0.5}$ samples with 2wt% of cometal additions show the best performance with their critical current density, $J_c$, and the pinning force, $F_p$, which might be attributable to providing effective flux pinning centres. Our study shows that the inclusion of a relatively small amount of Pb and Sn ($x = y \leq 0.02$) works effectively for the enhancement of superconducting properties with an improvement of intergrain connections as well as better phase uniformity.

## 1. Introduction

The Iron-based superconductors (FBS) have attracted significant attention owing to their relatively high superconducting transition temperature ($T_c$) of 58 K [1, 2, 3]. In 2008, FBS was discovered through F-doped LaOFeAs [3], and after that, more than 100 compounds have been reported under this high $T_c$ superconductors. On the basis of parent compound structure, these compounds can be categorized into 6-7 families [2, 4, 5, 6]: *RE*OFeAs (1111) (*RE* = rare earth), *A*Fe$_2$As$_2$ (*A* = Ba, K, Ca) (122), FeSe$_x$Te$_{1-x}$ (11), CaKFe$_4$As$_4$ (1144), and LiFeAs (111), 11 family. FeSe belongs to 11 family [7, 8] and has the simplest crystal structure [9, 10] in FBS. Generally, it shows the superconducting transition at 8 K which can be significantly enhanced up to 37.6 K under the applied external pressure of ~ 4.15 GPa [11]. Many new superconductors have been derived from FeSe with the enhanced superconductivity, including $A_x$Fe$_{2-y}$Se$_2$ (*A* = K, Rb, Tl, etc.) [12, 13] and other organic intercalated superconductors, (Li, Fe)OHFeSe [14], heavily electron-doped FeSe through gating or potassium deposition, and in particular, single-layer FeSe/SrTiO$_3$ films with a record high $T_c$ of ~100 K [15, 16].

Various kinds of doping have been reported such as Cu [17, 18], Ni [19], Cr [20], Co [8] at Fe sites and S [8, 21, 22], Te [8] at Se sites to understand the superconducting mechanism and to enhance the superconducting properties [23]. It has been reported that when

Te is substituted at Se-sites, the highest $T_c$ of up to 14.8 K is achieved with an optimal Te content of 50% [24]. Additionally, 11 family doesn't contain any dangerous or rare earth elements and shows a high critical density ($J_c \approx 8.6 \times 10^4$ A/cm$^2$ at 0 T, 2 K) and high upper critical field ($H_{c2} \approx 50$ T) [25, 26] for single crystals, which is hence interesting for a range of applications, such as superconducting magnets, wires, and tapes [27]. On the other hand, preparing single-phase superconducting bulks is difficult for this 11 family because the complicated phase diagram of FeSe has many stable crystalline forms such as tetragonal $\beta$-Fe$_x$Se, hexagonal $\delta$-Fe$_x$Se, orthorhombic FeSe$_2$, tetragonal $\beta$-Fe$_x$Se, monoclinic Fe$_3$Se$_4$, and hexagonal Fe$_7$Se$_8$, in which tetragonal phase generally exhibits the superconductivity with $T_c \sim 8$ K [7]. Some of these stable phases, particularly hexagonal $\delta$-Fe$_x$Se and hexagonal Fe$_7$Se$_8$, appear with the main tetragonal $\beta$-Fe$_x$Se phase during the growth process and are not suitable for superconducting properties [28, 29].

Several types of processes have been reported to enhance the flux pinning behaviours such as metal additions, chemical doping using different metallic and nonmetallic phases, high-energy irradiation, and the admixing of nanoparticles [30]. Recent studies have shown that different metal additions may be an effective and feasible approach for enhancing the superconducting properties of FBS by introducing additional pinning centres and comprehending the superconducting mechanism [30, 2, 6, 31, 32, 33]. In high $T_c$ cuprate superconductors, the critical current density $J_c$ of YBa$_2$Cu$_3$O$_y$ (YBCO) is enhanced by Ag addition [34]. Ag or Pb addition to 122 family (Sr$_{0.6}$K$_{0.4}$Fe$_2$As$_2$) also enhances the $J_c$ values with the improvement of grain connections [30]. In a similar way, various kinds of metal additions, such as Ag [35], Co [36], Ni [36], Li [37], Pb [38], Sn [39] are also reported for FeSe$_{0.5}$Te$_{0.5}$ bulks to enhance the superconducting properties. The reported studies suggest that the addition of Li, Pb, or Sn works as a positive effect to improve either transition temperature $T_c$ or critical current density $J_c$ [37, 38, 39]. Therefore, further research works are needed to fully understand the impact of adding suitable metal elements and their appropriate weight to bulk superconductors to enhance all of their superconducting properties,, *i.e.*, $T_c$ as well as $J_c$ of FBS, at the same time with high quality samples.

Chen *et al.* [39] reported 5 wt% ($x = 0$, $y = 0.05$) and 10 wt% ($x = 0$, $y = 0.10$) Sn added FeSe$_{0.5}$Te$_{0.5}$ samples where 5 wt% Sn added sample improved the superconducting offset transition temperature ($T_c^{offset}$) significantly by ~3 K compared to that of Sn-free sample but have almost the same onset transition temperature ($T_c^{onset}$) value as that of the parent compound. However, there is no report for a small amount of Sn addition, such as less than 5 wt% ($y<0.05$). Recently, Pb-added FeSe$_{0.5}$Te$_{0.5}$ has also been reported [38] and these results indicate that the superconducting transition is decreased and the impurity phase is enhanced with Pb addition due to the reduced Fe/Se/Te ratio from the stoichiometric FeSe$_{0.5}$Te$_{0.5}$ composition. However, 5 wt% Pb ($x = 0.05$, $y = 0$) addition has an onset $T_c$ of 13.8 K and improved the $J_c$ value in the measured magnetic field (up to 9 T) due to the improved grain connections. Hence, Pb addition weakens the superconducting transition of FeSe$_{0.5}$Te$_{0.5}$, while enhancing the intergranular behaviour and the critical current properties for a sample with a small amount of Pb ($x = 0.05$, $y = 0$). These reported studies suggest that it would be worthwhile to conduct additional research on the optimization of very low amounts of Pb and Sn addition, such as $x = y<0.05$, and processing parameters in order to improve superconductivity and critical current properties. However, there is no study available based on the co-metal addition to FeSe$_{0.5}$Te$_{0.5}$ polycrystalline samples or with other families of iron-based superconductors. Because Pb effectively increases the critical current density [38] and Sn improves the quality of the superconducting transition as reported [39], hence it would be interesting to investigate the superconducting properties of FeSe$_{0.5}$Te$_{0.5}$ with a small amount of both Pb and Sn addition, especially with a very low amount of additions. These are our main motivations behind this research paper.

In this study, we have synthesized a series of low amounts of Pb and Sn added FeSe$_{0.5}$Te$_{0.5}$ + $x$Pb + $y$Sn ($x = y = 0$, 0.01, 0.02, 0.03, 0.04, 0.05, and 0.10) and investigated the effects of Sn and Pb additions on the structure, microstructure, and superconducting properties of FeSe$_{0.5}$Te$_{0.5}$ bulks. Structural and microstructural analysis shows that the impurity phases are increased with higher Pb and Sn additions $x = y \geq 0.03$, however low amount of addition such $x = y \leq 0.02$ enhanced the superconducting transition by around 1 K and also improved the critical current density. Our present study shows that a small amount of cometal addition is an effective way to improve grain connectivity, superconducting transition $T_c$, and pinning behaviours, resulting an enhancement of the critical current density.

**Table 1.** The obtained lattice parameters '$a$' and '$c$', the impurity phases and crystallite size of the main tetragonal phase for FeSe$_{0.5}$Te$_{0.5}$ + $x$Pb + $y$Sn samples are listed. We have used Rigaku's PDXL software and the ICDD PDF4+ 2021 standard diffraction patterns database for the quantitative analysis of impurity phases (%) and crystallite size through the refinement of the measured XRD data.

| Sample | Lattice '$a$' (Å) | Lattice '$c$' (Å) | Pb$_{0.85}$Sn$_{0.15}$Te$_{0.85}$Se$_{0.15}$ (%) | Fe (%) | Hexagonal (%) | Crystallite size (nm) (FeSe$_{0.5}$Te$_{0.5}$ phase) |
|---|---|---|---|---|---|---|
| $x = y = 0$ | 3.7950 | 5.9713 | - | | 3 | 34 |
| $x = y = 0.01$ | 3.7977 | 5.9713 | ~2 | | < 2 | 48.2 |
| $x = y = 0.02$ | 3.7978 | 5.9665 | ~3 | | < 2 | 46.0 |
| $x = y = 0.03$ | 3.7958 | 5.9667 | ~6 | | -- | 42.1 |
| $x = y = 0.04$ | 3.7995 | 5.9684 | 9 | | -- | 35 |
| $x = y = 0.05$ | 3.7911 | 5.9639 | 14 | -- | -- | 33.5 |
| $x = 0.05, y = 0$ | 3.79309 | 5.9611 | -- | -- | -- | 45 |
| $x = y = 0.1$ | 3.7921 | 5.9681 | 29 | 2 | - | 24.5 |

## 2. Experimental details

The solid-state reaction method was used to grow the polycrystalline samples with nominal compositions of FeSe$_{0.5}$Te$_{0.5}$ + $x$Pb + $y$Sn ($x = y$ = 0, 0.01, 0.02, 0.03, 0.04, and 0.10). The initial steps were involved mixing the starting materials, which were Fe powder (99.99% purity, Alfa Aesar), Se (99.99% purity, Alfa Aesar), and Te (99.99% purity, Alfa Aesar), in accordance with the stoichiometric ratios of FeSe$_{0.5}$Te$_{0.5}$, for 15 minutes. More details about the synthesis process are reported elsewhere [38]. In the first step, the prepared pellets were sealed in an evacuated quartz tube which was heated at 600°C for 11 hours in a box furnace. In the second stage, the prepared pellets were ground and mixed with 1 wt% ($x = y = 0.01$), 2 wt% ($x = y = 0.02$), 3 wt% ($x = y = 0.03$), 4 wt% ($x = y = 0.04$), and 5 wt% ($x = y = 0.05$) and 10 wt% ($x = y = 0.1$) Pb and Sn (99% purity of Pb and Sn powder, respectively). These powders were pressed into pellets and sealed in an evacuated quartz tube, which was heated at 600°C for 4 h followed by furnace cooling process. The final pellets have a diameter of 12mm with 2.5 mm thickness. To reduce oxygen and moisture during the synthesis, we performed all of the growth process inside an inert gas glove box. Different batch samples were prepared to confirm the reproducibility of these bulk samples in terms of superconducting properties.

The structural analysis of all prepared samples were examined using the powder X-ray diffraction method (XRD), which was performed on a Rigaku SmartLab 3kW diffractometer with filtered Cu-K$\alpha$ radiation (wavelength: 1.5418 Å, power: 30 mA, 40 kV), and a Dtex250 linear detector. The measuring profile was used from 5° to 70° with a very small step of 0.01°/min. The measured XRD data were analysed using ICDD PDF4+ 2021 standard diffraction patterns database and Rigaku's PDXL software as well as Rietveld refinements using the Fullprof software [40] to perform the profile analysis, the quantitative values of impurity phases (%), and lattice parameter analysis for various samples. Microstructural characterisation was carried out using a field-emission scanning electron microscope. The magnetic measurements up to 9 T in the temperature range of 5–25 K under zero-field and field–cooling circumstances were carried out by Quantum Design PPMS using a vibrating sample magnetometer (VSM). During zero-field cooled (ZFC), the bulk sample was cooled down to 4 K, and then, after applying a magnetic field, the magnetic data was collected with increasing temperature of 5 to 25 K. A closed-cycle refrigerator was used to measure the temperature dependence of the resistivity of rectangular shaped samples in a zero magnetic field with various applied electric currents in a temperature range of 7 K to 300 K during the warming process.

## 3. Results and discussion

Powder X-ray diffraction patterns of $FeSe_{0.5}Te_{0.5}$ with various amounts of Pb and Sn additions ($FeSe_{0.5}Te_{0.5} + x$Pb $+ y$Sn) are depicted in Figure 1(a). All samples have shown the main tetragonal phase with space group *P4/nmm*. The parent compound ($x = y = 0$) has also shown a small amount (~3-4%) of hexagonal phase which is similar to that of the previous reported papers [41, 39, 38]. The diffracted peaks are not deviated by the additions of Pb and Sn, according to a comparison of the XRD patterns of the parent compound with Pb and Sn added samples, as shown in Figure 1(a). It suggests that Pb and Sn do not enter into the tetragonal structure of $FeSe_{0.5}Te_{0.5}$. We have also depicted the refined XRD patterns for low amounts of Pb and Sn added samples such as for $x = y = 0.01$, 0.02, and 0.03 in Figures 1(b), (c) and (d), respectively. The obtained lattice parameters and the qualitative values of the impurity phases for various samples are listed in Table 1. The superconducting phase's crystallite size, as estimated by the XRD fitting data, was also mentioned in Table 1. The crystal size was greater for the sample with $x = y = 0.01$ and 0.02 than that of other samples, but as further Pb and Sn additions, the crystal size shrank.

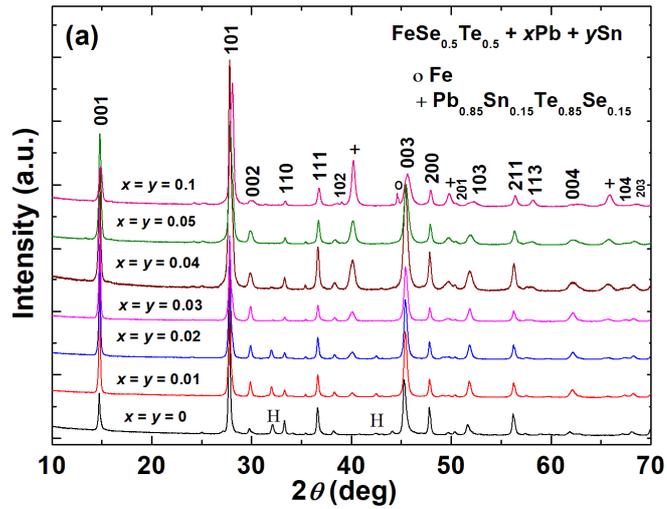

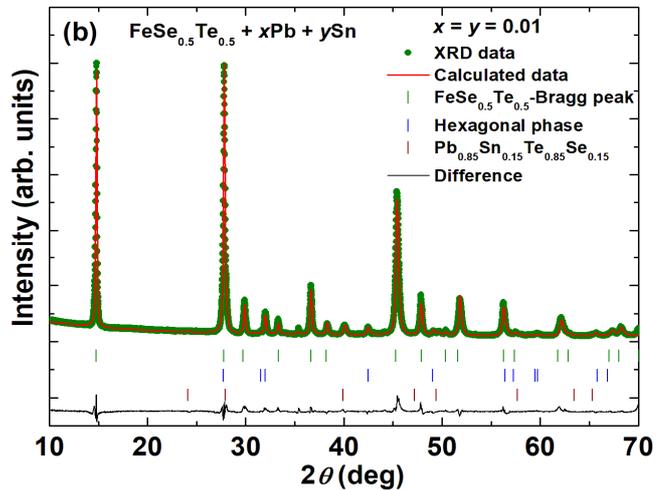

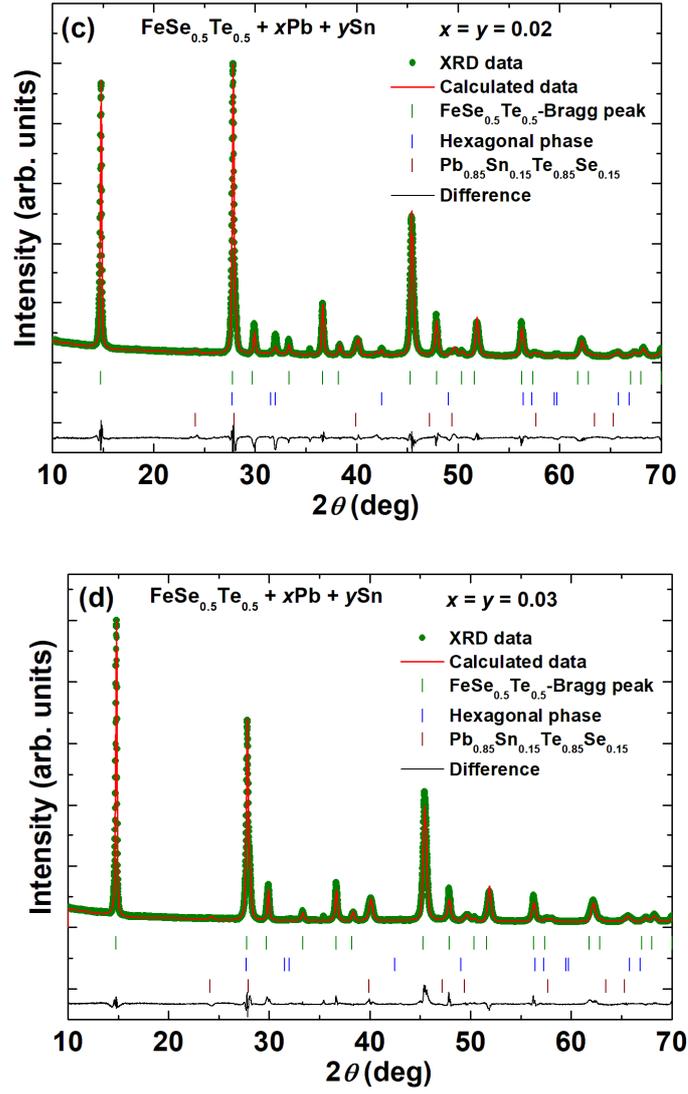

**Figure 1. (a)** X-ray diffraction patters (XRD) of powdered FeSe$_{0.5}$Te$_{0.5}$ + $x$Pb + $y$Sn ($x$ = 0, 0.01, 0.02, 0.03, 0.04, 0.05 and 0.1) samples. The fitted XRD patterns with the experimental, calculated diffraction patterns and their differences at the room temperature are shown for the sample with **(b)** $x = y = 0.01$, **(c)** $x = y = 0.02$ **(d)** $x = y = 0.03$. Instead of the nominal composition of FeSe$_{0.5}$Te$_{0.5}$, the tetragonal phase of Fe$_{1.1}$Se$_{0.5}$Te$_{0.5}$ was observed as the real composition of the superconducting phase. One hexagonal phase, Fe$_7$Se$_8$, was left out in the refinement because of relatively weak reflections, while a hexagonal phase of Fe$_{0.6}$Se$_{0.54}$Te$_{0.46}$ (~ 4–5%) has been found and is depicted as 'H' in figure (a). The list of the obtained lattice parameters '$a$' and '$c$', and the obtained phases are listed in table 1

The parent compound has the lattice parameters ($a$ = 3.79502 Å, $c$ = 5.9713 Å) which are almost same as the reported ones for bulk ($a$ = 3.7909 Å, $c$ = 5.9571 Å) and single crystals ($a$ = 3.815Å, $c$ = 6.069 Å) of FeSe$_{0.5}$Te$_{0.5}$ [41]. Interestingly, the hexagonal phase is notably reduced by a small amount of Pb and Sn addition ($x = y = 0.01$) and completely eliminated for $x = y = 0.03$ as depicted in Figure 1(a)-(d), and this phase is not seen even at higher Pb and Sn additions similar to those reported for Pb [38] or Sn additions [39]. However, for Pb and Sn added samples, Pb$_{0.85}$Sn$_{0.15}$Te$_{0.85}$Se$_{0.15}$ phase was appeared as an impurity phase which is very tiny for $x = y = 0.01$ and 0.02 but increases in intensity with further increase of Pb and Sn additions. Impurity phase enhancement is very similar to that of Sn or Pb added FeSe$_{0.5}$Te$_{0.5}$ [39, 38]. In the case of Sn-added FeSe$_{0.5}$Te$_{0.5}$ [39], SnSe$_{0.3}$Te$_{0.7}$ and Fe$_3$O$_4$ exist in the Sn-added samples, and their intensities increase as the amount of Sn addition increases. In higher Pb addition to bulk FeSe$_{0.5}$Te$_{0.5}$ [38], three extra phases such as PbTe,

FeSe$_{1-\delta}$, and Fe appeared as the impurity phases in which PbTe is observed a dominant impurity phase, suggesting a lower Te-contents in the FeSe$_{0.5}$Te$_{0.5}$ composition. The existence of Pb$_{0.85}$Sn$_{0.15}$Te$_{0.85}$Se$_{0.15}$ phase in this present study suggests the reduced concentration of Se/Te in the FeSe$_{0.5}$Te$_{0.5}$ composition. At high amounts of Pb and Sn additions, we have also observed a small amount of Fe as an impurity phase, as mentioned in table 1. The obtained lattice parameters for various samples, Table 1, indicate divergence with cometal additions with respect to the parent compound ($x = y = 0$) which suggests slightly lower Te/Fe/Se contents. Due to the presence of the various impurity phases, the refinement error is slightly higher for large amounts of Sn and Pb additions. It is important to note that excessive Sn and Pb additions can decrease the Fe/Te/Se concentrations in FeSe$_{0.5}$Te$_{0.5}$ compositions, while moderate levels of these additions can promote the formation of a tetragonal superconducting phase, similar to what has been observed in Pb or Sn-added FeSe$_{0.5}$Te$_{0.5}$ [39, 38].

These polycrystalline samples with $x = y = 0$, 0.01, 0.02, 0.03, 0.04, 0.05 and 0.1 were also subjected to an elemental analysis using the energy dispersive x-ray (EDAX) method, which allows for the measurement of the actual composition of the elements, as listed in Table 2. The homogenous distribution of the constituent elements are observed for $x = y = 0$ and 0.01, 0.02, and 0.03, as confirmed by results in Table 2. However, because of the impurity phase, the distribution of Sn and Pb in the samples with $x = y \geq 0.04$ is not uniform, and some regions were found to be rich in Pb, Sn, Se, and Te, which proposes the existence of an impurity phase of Pb$_{0.85}$Sn$_{0.15}$Te$_{0.85}$Se$_{0.15}$, consistent with XRD results. The parent compound shows the molar ratio of 1:0.49:0.51 which is almost the same as the low amount of cometal additions. However, with a high amount of Pb and Sn additions, deviation with this molar ratio and the actual weight percent of cometal increases. These findings suggest that excessive additions of Sn and Pb result in non-uniform element distributions.

**Table 2.** List of molar ratio of various elements presented in FeSe$_{0.5}$Te$_{0.5}$ + $x$Pb + $y$Sn bulks.

| Sample | Fe molar ratio | Te molar ratio | Se molar ratio | Pb (%) | Sn (%) |
|---|---|---|---|---|---|
| $x = y = 0$ | 1 | 0.49 | 0.5 | - | - |
| $x = y = 0.01$ | 1 | 0.5 | 0.49 | 0.98 | 0.99 |
| $x = y = 0.02$ | 1 | 0.5 | 0.5 | 1.5 | 2 |
| $x = y = 0.03$ | 1 | 0.52 | 0.48 | 2.4 | 2.6 |
| $x = y = 0.04$ | 1 | 0.53 | 0.42 | 3.16 | 4.1 |
| $x = y = 0.05$ | 1 | 0.51 | 0.58 | 3.7 | 5.1 |
| $x = 0.05, y = 0$ | 1 | 0.51 | 0.49 | - | - |
| $x = y = 0.1$ | 0.98 | 0.47 | 0.57 | 4.5 | 7.98 |

To perform the microstructural analysis, we have polished the pellet samples by using micron paper inside the glove box and collected backscattered scanning electron microscopy (BSE-SEM, revealing chemical contrast) images at different magnifications for different Sn and Pb-added samples. Figure 2 shows BSE images for $x = y = 0$, 0.02, 0.03 and 0.1 from low to high magnification images, respectively. We have observed three contrasts in our samples: light gray, white, and black contrasts corresponding to the phases of FeSe$_{0.5}$Te$_{0.5}$, Pb$_{0.85}$Sn$_{0.15}$Te$_{0.85}$Se$_{0.15}$, and pores, respectively. The parent compound has light gray and black contrasts that are observed almost homogeneous in microstructure images on microscale, as depicted in Figures 2(a)-(c). Furthermore, these images also confirm that the samples with $x = y = 0$ have many well-connected and disk-shaped grains with an average size of ~1–3 μm, and at some places, micropores are also observed. A minor amount of Sn and Pb addition ($x = y = 0.01, 0.02$) slightly increased the grain size (~3-4 μm) while decreasing pore sizes (from micro to nano range). Hence, many nanopores are observed, which results the improved grain connectivity and sample density due to the reduced pore size compared to the parent compound, as shown in Figure 2(d)-(f)). Furthermore, regarding the phase of $Pb_{0.85}Sn_{0.15}Te_{0.85}Se_{0.15}$, we have seen a few brighter contrasts in the sample (Figure 2(d)-(f)), as similar to XRD analysis.

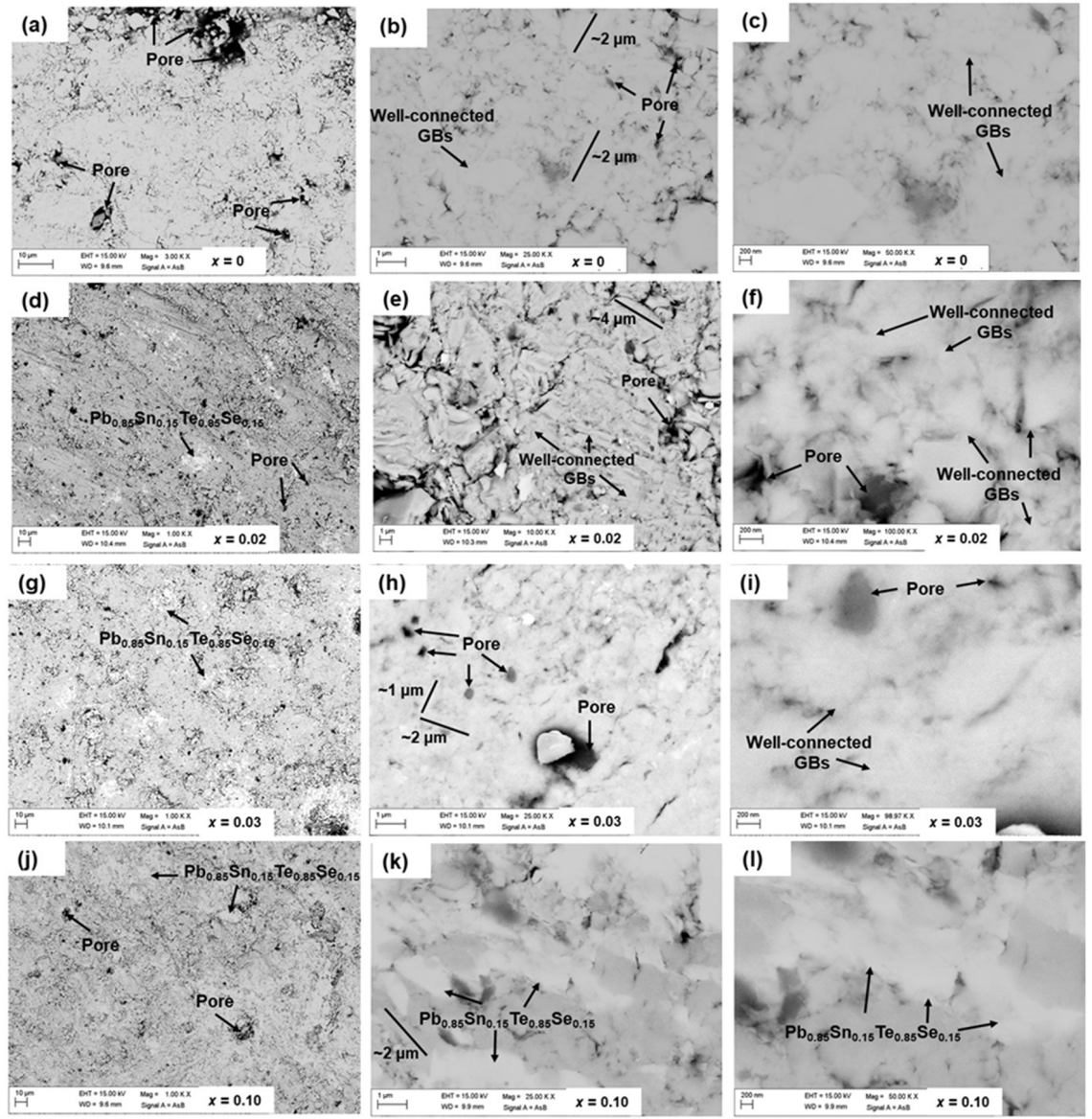

**Figure 2.** Back-scattered (BSE) images of various Pb and Sn-added $FeSe_{0.5}Te_{0.5} + x$Pb + $y$Sn polycrystalline samples: **(a)-(c)** for $x = y = 0$; **(d)-(f)** for $x = y = 0.02$; **(g)-(i)** for $x = y = 0.03$ and **(j)-(l)** for $x = y = 0.10$. Bright contrast, light gray and black contrast correspond to the phase of $Pb_{0.85}Sn_{0.15}Te_{0.85}Se_{0.15}$, Fe(Se,Te) and pores, respectively. The approximate grain size is depicted by the solid line.

With further increasing of Pb and Sn additions, the improvement of microstructure was observed with the enhancement of brighter phase with respect to $Pb_{0.85}Sn_{0.15}Te_{0.85}Se_{0.15}$, as shown in Figure 2(d)-(f). It seems that $Pb_{0.85}Sn_{0.15}Te_{0.85}Se_{0.15}$ phase filled up many nanopores, so we have observed comparatively fewer nanopores for $x = y = 0.02$ compared to the bulk samples with $x = y = 0.01$ but has almost the same grain size of ~3-4 μm. Hence, it suggests further improvement of grain connectivity and the density of materials. Figures 2(g)-(i) show BSE images for $x = y = 0.03$ where the most prominent phase of $Pb_{0.85}Sn_{0.15}Te_{0.85}Se_{0.15}$ is observed as a white contrast randomly in the bulk sample at many places, *i.e.* inside grains and at grain boundaries, and also the size of pores as a black contrast is increased compared to samples with low Pb and Sn additions ($x = y \leq 0.02$). The existence of pores and impurity phases in the samples results in the weak grain connections, and the plate-shaped grains are observed with an average grain-size of ~1-2 μm, as observed from Figure 2(g)-(i). For further cometal additions ($x = y$>0.03), white contrast ($Pb_{0.85}Sn_{0.15}Te_{0.85}Se_{0.15}$) is observed in larger areas and at many regions of the sample, and the reduced grain size is also observed as depicted in Figure 2(j)-(l) for $x = y = 0.1$. The increased impurity phase ($Pb_{0.85}Sn_{0.15}Te_{0.85}Se_{0.15}$) that is sandwiched between $FeSe_{0.5}Te_{0.5}$ grains often considerably reduces grain-to-grain connections and creates a strong barrier to intergranular supercurrent routes. Since it is well known from other iron-based superconductors that substantial cracking occasionally occurs at grain boundaries and within grains, but we have not seen any micro-cracks between the grains in our any bulk samples [42, 43]. Since $FeSe_{0.5}Te_{0.5}$ has a theoretical density of 6.99 g/cm$^3$ [44], on this basis, we have calculated the sample density by assuming the pure phase of $FeSe_{0.5}Te_{0.5}$ for our various samples, which are obtained around 51%, 61.9%, 65.6% and 50.8%, for $x = y = 0$, 0.01, 0.02, and 0.03, respectively. It indicates that a very small amount of Sn and Pb content added to the parent sample slightly enhanced the sample density as also observed from the microstructure analysis. Analysis of Figure 2 clearly demonstrates that a very small amount of Pb and Sn addition ($x = y$≤0.02) improves grain connectivity, and sample density, and decreases pores in contrast to a larger amount of Pb and Sn additions ($x = y$≥0.02), which reduce phase purity and cleanness of grain boundaries and increases the number of pores. Non-superconducting phases at the grain boundaries of $FeSe_{0.5}Te_{0.5}$ for higher cometal additions generally create a problem for superconducting properties, as also reported for Pb-added Sr122 [45], and Pb-added $FeSe_{0.5}Te_{0.5}$ [38], and Sn-added $FeSe_{0.5}Te_{0.5}$ [39]. As a result, our analysis suggests that a very small amount of Pb and Sn additions work effectively to increase material density while also improving grain size and connectivity.

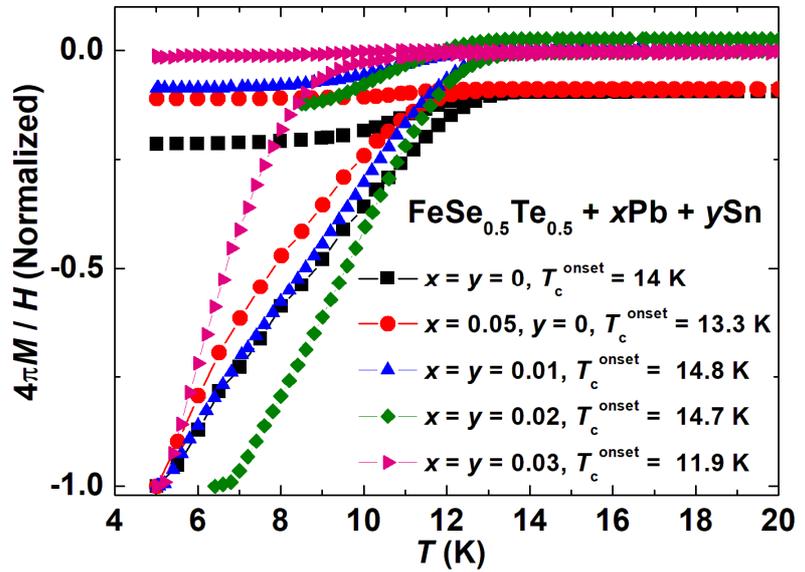

**Figure 3.** The variation of magnetic susceptibility ($\chi = 4\pi M / H$) with temperature for various $FeSe_{0.5}Te_{0.5} + xPb + ySn$ ($x = y = 0$, 0.01, 0.02, 0.03 and also $x = 0.05$, $y = 0$) bulks at the applied magnetic field of 20 Oe under zero field-cooled (ZFC) and field-cooled (FC) regimes.

Figures 3 depicts the DC magnetic susceptibility ($\chi = 4\pi M/H$) in both zero-field-cooled (ZFC) and field-cooled (FC) magnetization curves for samples, $x = y = 0$ and $x = 0.05$, $y = 0$; $x = y = 0.01$, $x = y = 0.02$ and $x = y = 0.03$ measured under an applied magnetic field of 20 Oe in the temperature range 5-20 K. We have shown the normalized magnetic susceptibility for all these samples for a comparison point of view. What one can safely conclude from Figure 3 is that the studied samples are a bulk superconductor. Superconducting transition is observed at 14 K with a sharp diamagnetic transition in the magnetic susceptibility ($\chi$) in both ZFC and FC situations for the parent compound ($x = y = 0$). Only

Pb-added sample ($x = 0.05$, $y = 0$) shows the onset transition at 13.3 K and has a broader transition than that of the parent compound. Interestingly, a small amount of Pb and Sn such as $x = y = 0.01$ slightly enhanced the transition temperature ($T_c$ ~14.8 K) with the sharpness of transition compared to the sample $x = 0.05$, $y = 0$. Further addition of Pb and Sn, almost the same superconducting onset transition of 14.7 K is observed for $x = y = 0.02$ with better sharpness of the transition compared to other samples. However, further increase of Pb and Sn additions reduce the transition temperature with large broadening of the transition. It might be possible due to the formation of impurity phase $Pb_{0.85}Sn_{0.15}Te_{0.85}Se_{0.15}$ and to reduce the actual content of Te and Se from the main phase $FeSe_{0.5}Te_{0.5}$ as discussed above with XRD data and microstructural analysis. Single step transition of each samples can be explained by the intergranular properties of these bulk samples, as discussed and reported for other FBS families [46]. These analyses also confirm that a very low amount of Sn and Pb added samples ($x = y \leq 0.02$) are effective for the superconducting properties of $FeSe_{0.5}Te_{0.5}$ as similar to the conclusion of microstructural analysis and XRD measurements. Further $T_c$ is decreased as Sn and Pb concentrations increased, possibly due to changes in Te/Se concentrations.

The temperature dependence of the resistivity ($\rho$) are shown in Figures 4(a)–(c) for the nominal compositions of polycrystalline $FeSe_{0.5}Te_{0.5} + xPb + ySn$ ($x = y = 0$–0.1) in zero magnetic field. Due to the structural phase transition, the parent $FeSe_{0.5}Te_{0.5}$ ($x = y = 0$) exhibits a large anomaly in resistivity at a temperature of below ~110 K [47]. As reported [38] for Pb added Fe(Se,Te), the electrical behaviour of this sample is gradually changed and a somewhat higher value of the normal state resistivity was observed due to the tiny amount and uneven distribution of the impurity PbTe phase with only Pb addition ($x = 0.05$, $y = 0$), and this resistivity anomaly also appeared for this Pb added samples ($x = 0.05$, $y = 0$) [38]. Small amount of Sn and Pb addition to $FeSe_{0.5}Te_{0.5}$ up to $x = y = 0.03$ increases the metallic behaviour and its resistivity decreases in the whole temperature range. Interestingly, the anomaly related to the structural phase transition is also observed for these samples. A kink or concavity feature is appeared for samples with very low amounts of Pb and Sn ($x = y = 0.01$, 0.02 and 0.03) below 80 K, which is similar to the behavior reported for FeSe [48] or Fe(Se,Te) samples [49], and is usually linked with the weak structural distortion or attributed to the weak localization effect [48, 49].

Further enhancement of Pb and Sn additions ($x = y \geq 0.03$), the resistivity start to increase in the normal state and showed semi-metallic behaviour below the structural phase transition. The amount of $Pb_{0.85}Sn_{0.15}Te_{0.85}Se_{0.15}$ phase is enhanced very rapidly for samples with $x = y > 0.03$ as discussed above and its distribution inside the sample becomes more homogeneous with reduced whole sample density as observed from the microstructural analysis, which could be a reason for the enhancement of the normal state resistivity, as in Figure 4(a), which is visible more clearly below the structural transition. The sample with $x = y = 0.1$ has shown high resistivity values within the whole measured temperature range due to very large amount of impurity phases. However, the low amount of addition of Pb and Sn ($x = y \leq 0.02$) increases the density of samples, as discussed in the microstructural analysis, which could be a reason for the decreased resistivity of these samples and supports the formation of the superconducting tetragonal phase. The observed properties of the sample with $x = y = 0.03$ depict the combined effect of low and high amounts of Pb and Sn additions, suggesting that it could be an optimum cometal addition level. Due to the presence of impurity phases, the higher Sn and Pb-added sample ($x = y \geq 0.02$) has a negative slope of resistivity below 120 K, which primarily manifests as semi-metallic behaviour.

Low temperature behaviour of the resistivity ($\rho$) as a function of temperature from 5 K to 18 K is shown in Figure 4(b) where each sample depict the a superconducting transition. The parent compound shows the transition temperature around 14.8 K with transition width ($\Delta T$) of 3.1 K. The sample with $x = y = 0.01$ and 0.02 have an enhanced transition temperature of 15.6 K and 15.4 K respectively with sharper superconducting transition. Further increased of Sn and Pb additions, the transition temperature is decreased with the broader transition width. Interestingly, the sample with $x = y = 0.03$ shows the onset transition of 12.8 K. Further increase of Sn and Pb additions, the onset $T_c$ reduces very slowly but exhibits a relatively broad transition with a low $T_c^{offset}$. The onset $T_c$ is observed around 12.1 K, 11.87 K and 11.6 K for the samples with $x = y = 0.04$, $x = y = 0.05$ and $x = y = 0.1$, respectively.

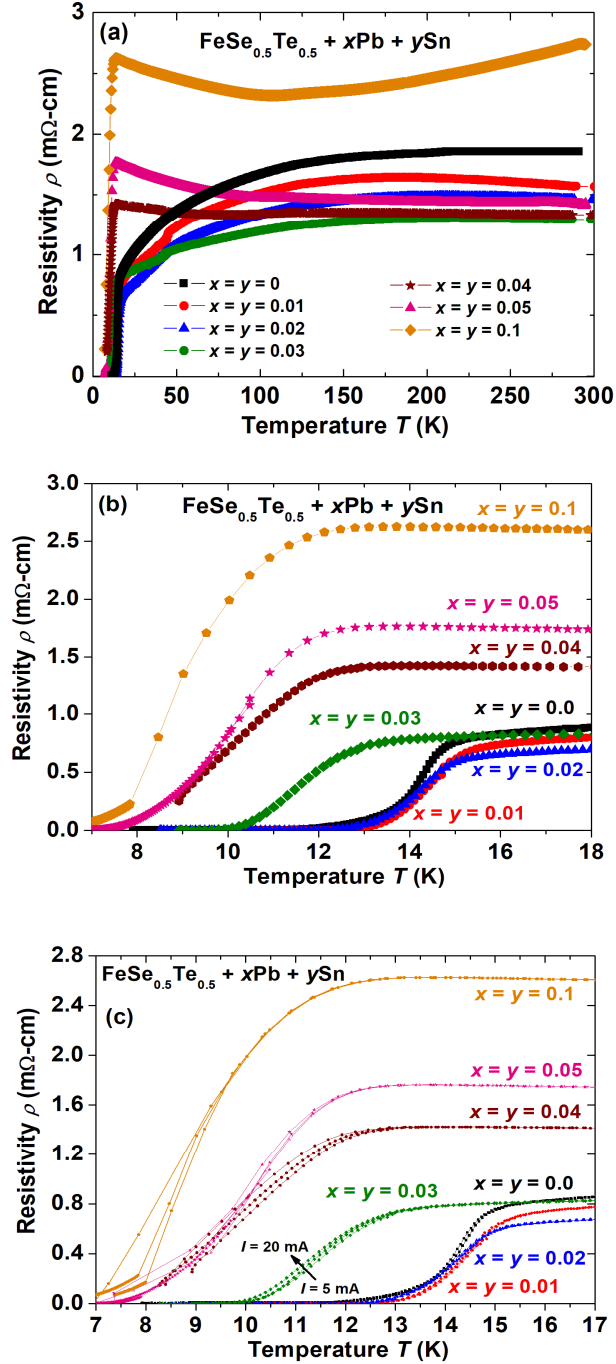

**Figure 4. (a)** The variation of resistivity ($\rho$) with respect to the temperature for all Pb and Sn added FeSe$_{0.5}$Te$_{0.5}$ bulks (FeSe$_{0.5}$Te$_{0.5}$ + $x$Pb + $y$Sn ($x = y = 0$, 0.01, 0.02, 0.03, 0.04, 0.05 and 0.1)). **(b)** The resistivity behaviours with temperature for various samples in low temperature region (< 20 K). **(c)** The low temperature resistivity variation with temperature for FeSe$_{0.5}$Te$_{0.5}$ + $x$Pb + $y$Sn ($x = y = 0$, 0.01, 0.02, 0.03, 0.04, 0.05 and 0.1) with respect to different currents $I$ = 5, 10, 20 mA.

More interestingly, their $T_c^{offset}$ values differ significantly. According to reports, samples with 5% Pb addition show comparable $T_c^{onset}$ values (13.8 K), which is around 1.1 K lower than the value for the Pb-free sample [38]. Chen et al. [39] has reported that 5% Sn added FeSe$_{0.5}$Te$_{0.5}$ ($x$ = 0, $y$ = 0.05) has $T_c^{onset}$ = 13.8 K and $T_c^{offset}$ = 12 K with respect to $T_c^{onset}$ = 13.5 K and $T_c^{offset}$ = 9 K of the parent compound which has dramatically enhanced the zero resistivity temperature ($T_c^{offset}$) by 3K accompanied by almost same onset temperature of superconducting transition ($T_c^{onset}$) [39]. Interestingly, low amount of cometal Sn and Pb addition improved the onset transition temperature and also reduced the transition width which work well accordingly to previous studies [39, 38]. The sharper transition for 1 and 2wt% Sn and Pb added sample ($x$ = $y$ = 0.01, 0.02) suggests better grain connections and slightly higher Te/Se concertation than the Sn and Pb-free one ($x$ = $y$ = 0) which might be due to reducing the hexagonal phase, as discussed for XRD measurements. On the other hand, further increment of Sn and Pb addition exhibits the broadening of superconducting transition which might result due to the increased impurity phase (Pb$_{0.85}$Sn$_{0.15}$Te$_{0.85}$Se$_{0.15}$) and the decreased superconducting phase. The slight decrease in the lattice parameters with Sn and Pb addition, as mentioned in Table 1, suggests that there is less Se/Te concentration in FeSe$_{0.5}$Te$_{0.5}$ composition, which is also supported by EDAX measurements (Table 2). This could be a possible reason for the reduced transition temperature $T_c$ at high Pb and Sn additions. The reported study based on Li doped FeSe$_{0.5}$Te$_{0.5}$ [37] has confirmed that the doping element Li enter the crystal structure of Fe(Se,Te) and enhanced the superconducting transition by 1-1.5 K for 1 wt% doping without affecting the $T_c^{offset}$. In contrast to these earlier findings, adding 5% Sn to FeSe$_{0.5}$Te$_{0.5}$ can significantly raise $T_c^{offset}$ by 3 K without affecting $T_c^{onset}$ while not altering the crystal structure of the compound. The magnetic elements such as Co and Ni at Fe sites reduce the superconducting properties of FeSe$_{0.5}$Te$_{0.5}$ [41]. Our current results show the enhancement of $T_c^{onset}$ by ~1 K and also slightly improved $T_c^{offset}$ by very small amount of Sn and Pb added samples without entering the crystal structure of FeSe$_{0.5}$Te$_{0.5}$ which implies that small amount of Sn and Pb ($x$ = $y$ ≤ 0.02) seems to be the most promising additive among metals to further improve the superconductivity in the 11-type FBS.

Since the offset transition temperature ($T_c^{offset}$) generally relates to the grain connections *i.e.* the intergrain effect, whereas the onset transition temperature ($T_c^{onset}$) represents the specific grain effect *i.e.* intragrain effect [50, 51]. These affects can be understand by the resistivity measurements under different applied currents. To understand the grain connectivity behaviours of our bulk samples, we have depicted the low temperature resistivity behaviours of various bulk samples with three different currents $I$ = 5, 10, and 20 mA in Figure 4(c). The bulk samples with $x$ = $y$ = 0.01 and 0.02 have almost no transition broadening with various currents and also a sharper transition compared to that of the parent compound ($x$ = $y$ = 0). The transition broadening is increased for higher Pb and Sn additions ($x$ = $y$≥0.02), and the offset transition is more sensitive with the applied currents, as shown in Figure 4(c) which could be due to the enhanced impurity phases as observed from XRD patterns. It clearly suggests that low amount of co-metal ($x$ = $y$ = 0.01 and 0.02) added samples have a better integrain effect than that of the parent compound. These outcomes support the analysis of microstructural studies, as discussed above. Previous study [38] shows that 5 wt% Pb added FeSe$_{0.5}$Te$_{0.5}$ has almost the same broadening with applied current as that of the parent compound but have a shaper transition. However, higher Pb additions reduce the grain connections due to the enhancement of impurity phase. Compared to our results with only Pb added samples, low amount of co-metal additions to FeSe$_{0.5}$Te$_{0.5}$ has almost no broadening of the transition with respect to the applied current which suggests better grain connectivity. These results are well agreed with microstructural and XRD analysis.

Magnetic moment hysteresis loops *M(H)* at constant temperature 7 K for $x$ = $y$ =0, $x$ = 0.05, $y$ = 0; $x$ = $y$ = 0.01, 0.02 and 0.03 were measured with the rectangular-shaped sample in order to determine the persistent critical current density $J_c$. The measured magnetic loops *M(H)* for these samples were observed under the ferromagnetic effects, which is similar to the previous reports based on FeSe samples [38, 29, 52]. The inset of Figure 5(a) shows *M(H)* loop for Pb and Sn added samples with $x$ = $y$ = 0.02 which is depicted after the subtraction of the normal state magnetization *i.e M(H)* loop at 22 K. Similar magnetization loops, however, with larger backgrounds, have been obtained for a sample with high Pb and Sn additions.

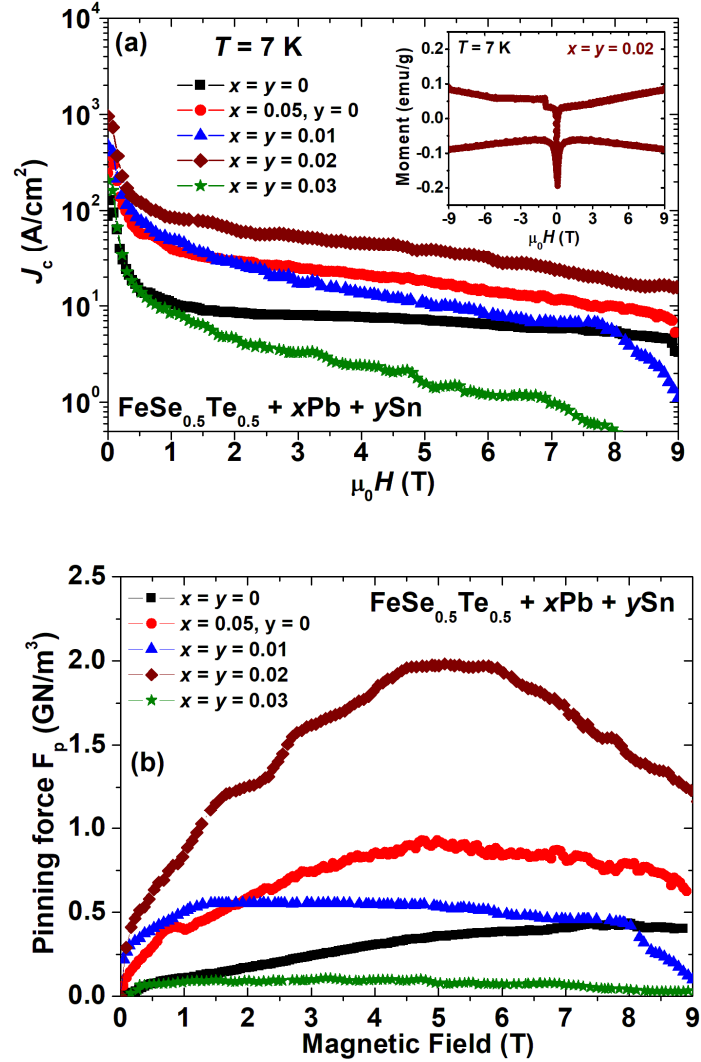

**Figure 5. (a)** The magnetic field ($H$) dependence of critical current density ($J_c$) for FeSe$_{0.5}$Te$_{0.5}$ + $x$Pb + $y$Sn ($x = y = 0, 0.01, 0.02, 0.03$ and $x = 0.05, y = 0$)) samples with respect to the parent compound FeSe$_{0.5}$Te$_{0.5}$ at temperature of 7 K. The inset figure shows the magnetic hysteresis loop $M(H)$ at 7 K for $x = y = 0.02$ after the subtraction of the normal state background. **(b)** The variation of pinning force $F_p$ with respect to the applied magnetic field at 7 K for various bulk FeSe$_{0.5}$Te$_{0.5}$ + $x$Pb + $y$Sn ($x = y = 0, 0.01, 0.02, 0.03$ and $x = 0.05, y = 0$)) samples.

These hysteresis loops allow us to estimate the critical current density, which is an important parameter for practical applications. The Bean critical state model [53] was applied to obtain the critical state densities from the magnetization loops. The calculation of the critical current density $J_c$ for our samples was performed using the formula $J_c = 20\Delta m/Va(1-a/3b)$ [53], where $\Delta m$ is the hysteresis loop width, $V$ is the volume of the sample, and $a$ and $b$ are the lengths of the shorter and longer edge, respectively. Figure 5(a) depicts the magnetic field dependence of the critical current density ($J_c$) up to 9 T at 7 K for the parent compound with various Pb and Sn added samples. $J_c$ values of the parent compounds were enhanced by adding 5-wt% Pb to FeSe$_{0.5}$Te$_{0.5}$ ($x = 0.05, y = 0$) whereas, with the addition of Sn and Pb, i.e. $x = y = 0.02$, $J_c$ values are further enhanced in the whole magnetic field range up to 9 T. Interestingly, the calculated $J_c$ of samples $x = y = 0.01$ and 0.02 has field dependence almost similar to that of Pb added samples ($x = 0.05, y = 0$) and enhanced one order of magnitude of the $J_c$ values compared to the parent compound. This improvement in $J_c$ values suggests that cometal inclusion is capable of providing effective flux pinning centres. It could be possible due to the increased density and improved grain connections caused by the addition of a small amount of Sn and Pb, which are clearly observed in the microstructural analysis and resistivity studies. In pure bulk $MgB_2$ polycrystalline samples, the same observation has been observed [54], where Ag nanoparticle addition enhances $J_c$ value due to extra pinning centres. One should note an important point that 5% Pb added sample ($x = 0.05$, $y = 0$) has almost the same $J_c$ values [38] and similar behaviour as that of 1% Sn and Pb added samples ($x = y = 0.01$). It clearly suggests that Sn can be the most effective metal to enhance the $J_c$ value for $FeSe_{0.5}Te_{0.5}$ samples, which is comparable to the reported elevation of $J_c$ values for Sn-added SmFeAs(O,F) [50] where Sn additions also work more effectively to improve the intergranular current than that of other metal additions [30].

To understand the pinning behaviours of these samples, the magnetic field dependence of vortex pinning force density, $F_p$, has been calculated by $F_p = \mu_0 H \times J_c$ [55] with the obtained $J_c$ values at 7 K which is depicted in Figure 5(b) for various samples. The $F_p$ curves of the parent compound increase with magnetic fields and reach a maximum around 8-9 T whereas 1-wt% Pb and Sn additions show a maximum of $F_p$ for low magnetic fields, and then it decreases very slowly with the applied field. Further Sn and Pb additions enhance $F_p$ values in the whole measured fields and shifted the maximum of $F_p$ to the higher magnetic field as similar to 5-wt% Pb added samples ($x = 0.05$, $y = 0$). The samples with $x = y = 0.03$ has showed the similar behaviours of 1-wt% Pb and Sn added $FeSe_{0.5}Te_{0.5}$ but with lower values of $F_p$ compared to all other samples depicted in Figure 5(b). This is an unusual behaviour, most likely caused by cometal addition, that warrants further investigation to understand how cometal additions can influence the vortex pinning mechanisms in $FeSe_{0.5}Te_{0.5}$ compounds. The $F_p$ values are enhanced up to the intermediate fields (~5-6 T) range for the small amount of Pb and Sn-added $FeSe_{0.5}Te_{0.5}$ ($x = y \leq 0.02$) compared to that of the parent compound ($x = y = 0$) which is in a nice agreement with the $J_c$ enhancement as depicted in Figure 5(a). 5-wt% Pb added samples ($x = 0.05$, $y = 0$) also enhanced the $F_p$ values, which are similar to the previous report [38], and higher than those of 1-wt% Pb and Sn added samples and the parent compounds. Furthermore, the obtained $F_p$ values of the parent compounds are almost the same as those reported (0.1-1 $GN/m^3$) in previous studies [56, 52] based on the polycrystalline Fe(Se,Te) samples. The $F_p$ behaviour leads us to the conclusion that improving the appropriate pinning centers is a reason for the enhancement of critical current behaviors. There are also reports of similar results for Ag-added $MgB_2$ [54] and Sn-added other FBS bulk samples [50]. High-pressure techniques such as high-pressure growth and high-pressure sintering can be used to further improve the $J_c$ and $F_p$ of these samples [2, 56].

To summarize the main findings of our study, the variation of transition temperature $T_c^{onset}$, the transition width ($\Delta T$), room temperature resistivity ($\rho_{300K}$) and $RRR$ ($\rho_{300K}/\rho_{20K}$) and the critical current density ($J_c$) for 0 T and 5 T at 7 K with weight concentration of Pb and Sn added samples ($x$, $y$) are shown in Figure 6(a)-(e). The $T_c^{onset}$ is enhanced by ~1 K for 1 and 2 wt% of Pb and Sn-added samples. With further increase of weight of this concentration, the $T_c^{onset}$ value starts to decrease (Figure 6(a)). The value of transition width $\Delta T$ (= $T_c^{onset} - T_c^{offset}$) also reduces with a small amount of Sn and Pb addition and reaches to a minimum value for 2% weight Pb and Sn addition i.e. it has a sharp transition with respect to other samples as depicted in Figure 6(b). This is a clear indication of higher homogeneity, better grain connectivity and a phase purity of this sample compared to other samples. On the other hand, the broadening of the transition i.e. the transition width $\Delta T$ starts to enhance with further increase of Pb and Sn additions and becomes almost saturated for $x = y \geq 0.04$. Addition of Pb and Sn also enhanced the metallic nature of $FeSe_{0.5}Te_{0.5}$ sample at room temperature i.e. the resistivity $\rho_{300K}$ decreases for low amount of metal additions, as shown in Figure 6(c), and $\rho_{300K}$ reaches to minimum values for 3 and 4%weight Sn and Pb added samples. Further enhancement of Pb and Sn, $\rho_{300K}$ is started to increase which is due to the enhancement of impurity phases as discussed above. We have also calculated and plotted the residual resistivity ratio $RRR$ value for all samples as depicted in Figure 6(d). The maximum *RRR* is observed for the samples with $x = y = 0.02$ and after that *RRR* has started to decrease with further increase of Pb and Sn additions.

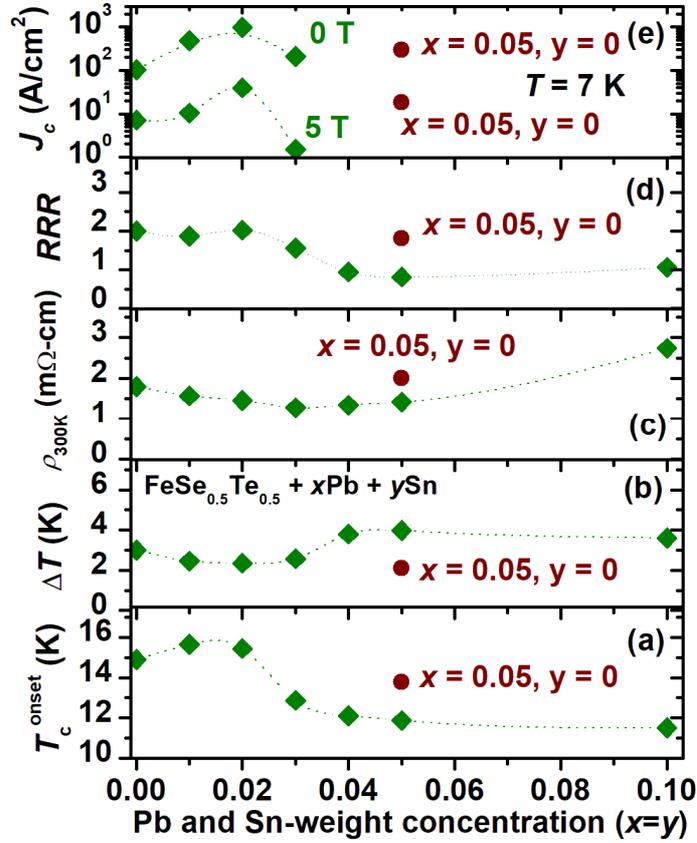

**Figure 6.** The variation of **(a)** transition temperature ($T_c$), **(b)** transition width ($\Delta T$), **(c)** room temperature resistivity $\rho_{300K}$ and **(d)** residual resistivity ratio *RRR* ($\rho_{300K}/\rho_{20K}$) **(e)** the critical current density $J_c$ for 0 T and 5 T at 7 K with respect to weight% of Pb and Sn added for parent $FeSe_{0.5}Te_{0.5}$ i.e. $FeSe_{0.5}Te_{0.5} + x$Pb $+ y$Sn ($x = y = 0, 0.01, 0.02, 0.03, 0.04, 0.05$ and $0.1$) and also only Pb added $FeSe_{0.5}Te_{0.5}$ ($x = 0.05, y = 0$).

The maximum of *RRR* and the minimum of $\Delta T$ are another transport signature of the high quality of the polycrystalline samples with $x = y = 0.02$. The onset $T_c$ is reduced, and the transition width, $\Delta T$ and $\rho_{300K}$ are enhanced with increasing Pb and Sn additions. It is worth noting that *RRR* for our best samples is 2.2, which is higher than the reported value (1.3) for 5 and 10% Sn added $FeSe_{0.5}Te_{0.5}$ samples [39] and also better than the reported (1.8) for Pb added $FeSe_{0.5}Te_{0.5}$ samples [38]. Very small amount of Sn and Pb-additions improves the overall *RRR* of the parent compound, as similar to those reported for Ag, Sn and Pb additions [32, 38, 39]. Meanwhile, 1 and 2 wt% Sn and Pb-added samples show ~1 K higher transition and a comparatively sharper transition width with $T_c$ of 15.6 K and $T_c^{\text{offset}}$ of 13.2 K. The transition width of 2.4 K suggests a sharper transition than for the pure sample. In Figure 5(e), we have plotted the $J_c$ value at 0 T and 5 T for various Pb and Sn added samples with parent and only 5% Pb added samples. It clearly indicates that a very small amount of addition of Sn and Pb acts as effective pinning centres and, in consequence, improves the critical current density by an order of magnitude with respect to the parent compound and also only Pb-added samples. This analysis suggests that a small amount of cometal addition improves both the superconducting properties and also the granular behaviour.

Disorder can significantly enhance the superconductivity and has been utilized as an effective way to explore superconducting order [57, 58, 59]. Strong disorder, on the other hand, increases phase fluctuations, which lowers the superfluid density and suppresses superconductivity globally [59, 60]. As the disorder strength is varied, an optimal degree of inhomogeneity can be reached which enhances the superconducting properties and the transition temperature $T_c$ reach to the maximum value. Outside that region, strong disorder reduces superconductivity and can even cause the superconductor-insulator transition [61] as observed in conventional superconductors, which are usually believed to be insensitive to small concentrations of random nonmagnetic impurities [62]. On this basis, here, we can explain the enhancement of superconducting properties of $FeSe_{0.5}Te_{0.5}$ with the correlation effect in the disorder which is generated by nonmagnetic cometal addition. A large amount of Pb and Sn addition generally creates a strong disorder due to a large amount of impurity phase, as discussed above and its behaviour shifted to superconductor-insulation transition which is clearly observed through the resistivity measurements for high Pb and Sn added $FeSe_{0.5}Te_{0.5}$ samples ($x = y \geq 0.04$) (Figure 4(a)). A very small amount of cometal addition such as (below $x = y = 0.01$) does not affect the superconducting properties, as observed from Figure 6 with dotted line, and more than 3 wt% cometal addition induces strong disorder which enhances rapidly with further Pb and Sn additions. On these analyses, we can conclude that 1 to 2 wt% cometal addition is the optimum region where the disorder strength improves the superconducting properties of $FeSe_{0.5}Te_{0.5}$ bulk. Hence, it seems that the enhanced superconductivity of these materials is related to the effects of the disorder correlations as well reported for other superconductors [59].

## 4. Conclusions

We have studied the cometal addition effect on the superconducting properties of $FeSe_{0.5}Te_{0.5}$ through the various characterizations. Structural analysis of the prepared $FeSe_{0.5}Te_{0.5}$ samples with Pb and Sn additions showed that these metals do not enter in superconducting tetragonal structure of $FeSe_{0.5}Te_{0.5}$ and lattice parameters seem to be unaffected by these additions. A large amount of Sn and Pb additions ($x = y > 0.02$) enhanced the impurity phases and introduced inhomogeneities into the samples, resulting a change in Fe/Se/Te ratio from the stoichiometric $FeSe_{0.5}Te_{0.5}$ composition. However, a very low amount of Pb and Sn additions were effective to enhance the transition temperature $T_c$ and $J_c$ in the measured magnetic field (up to 9 T) due to the improved grain connections as well as the presence of additional pinning centres. Microstructural analysis shows a disc shaped superconducting grains, and at high Pb and Sn additions, the intergrain connections are reduced compared to low amount of Sn and Pb added samples and the parent compounds. A cometal addition effect on iron-based superconductors has been studied for the first time, confirming cometal additions can be a potential way to enhance superconducting properties with the improvement of sample qualities. We believe this method will enable further exploration of Fe(Se,Te) and other FBS materials to achieve additional improvements in the superconducting properties and the development of their magnetic applications, especially with respect to superconducting wires and tapes.

**Funding:** This research was funded by NATIONAL SCIENCE CENTRE (NCN), POLAND, grant number "2021/42/E/ST5/00262" (SONATA-BIS 11). SJS acknowledges financial support from National Science Centre (NCN), Poland through research Project number: 2021/42/E/ST5/00262.